\documentclass[english]{article}
\usepackage[T1]{fontenc}
\usepackage[latin9]{inputenc}
\usepackage{geometry}
\usepackage{amsmath}
\usepackage{setspace}
\usepackage{amssymb}

\makeatletter
\newcommand{\lyxaddress}[1]{
\par {\raggedright #1
\vspace{1.4em}
\noindent\par}
}

\usepackage{babel}
\makeatother

\begin{document}

\title{SHAPE INVARIANT POTENTIAL AND SEMI - UNITARY TRANSFORMATIONS (SUT)
FOR SUPERSYMMETRY HARMONIC OSCILLATOR IN $T^{4}$- SPACE}

\author{P. S. Bisht and O. P. S. Negi}

\maketitle

\lyxaddress{\begin{center}
Department of Physics\\
Kumaun University\\
S. S. J. Campus\\
Almora - 263601 (India)
\par\end{center}}

\lyxaddress{\begin{center}
Email: - ps\_bisht123@rediffmail.com\\
ops\_negi@yahoo.co.in
\par\end{center}}

\begin{abstract}
Constructing the Semi - Unitary Transformation (SUT) to obtain the
supersymmetric partner Hamiltonians for a one dimensional harmonic
oscillator, it has been shown that under this transformation the supersymmetric
partner loses its ground state in $T^{4}$- space while its eigen
functions constitute a complete orthonormal basis in a subspace of
full Hilbert space.


\end{abstract}

There are a number of analytically solvable problems in non - relativistic
quantum mechanics for which all the energy eigen values and eigen
functions are explicably known. The question naturally arises as to
why these potentials are solvable and what is the underlying symmetry
property? No definite answer was known until 1983 when in a largely
unnoticed paper, Grenden Shtein \cite{key-1} pointed out that all
these potentials have a property of shape invariance. Keeping in view
the recent potential importance of tachyons \cite{key-2,key-3,key-4,key-5}
and the fact that these particles are not contradictory to special
theory of relativity and are localized in time in view of second quantization
and interaction of superluminal electromagnetic fields \cite{key-6,key-7},
we have constructed a Semi - Unitary Transformation (SUT) to obtain
the supersymmetric partner Hamiltonians for one dimensional harmonic
oscillator in $T^{4}$ - space (i.e the localization space for tachyons
\cite{key-8,key-9}) and it has been demonstrated that under this
SUT the supersymmetric partner $H_{+}^{T}$ loses its ground state
while its eigen functions constitute a complete orthonormal set in
a subspace of full Hilbert space.

In order to over come the various problems associated with superluminal
Lorentz transformations (SLTs) \cite{key-4,key-5} , six - dimensional
formalism \cite{key-10} of space time is adopted with the symmetric
structure of space and time having three space and three time components
of a six dimensional space time vector. In this formalism, a subluminal
observer $\mathcal{O}$ in the usual $R^{4}\equiv(\overrightarrow{r},\, t)$
space is surrounded by a neighborhood in which one measures the scalar
time $|t\,|\equiv|(t_{x}^{2}+t_{y}^{2}+t_{z}^{2})|^{\frac{1}{2}}$
and spatial vector $\overrightarrow{r}=(x,\, y,\, z)$ out of six
independent coordinates $(x,\, y\,,z\,,t_{x\,},t_{y}\,,t_{z}\,)$
of the six - dimensional space $R^{6}$. On passing from $R^{6}=(\overrightarrow{r},\overrightarrow{\, t})$
to $(R^{6})'=(\overrightarrow{r'},\,\overrightarrow{t}')$ via SLT's,
the usual $(R^{4})'=(\overrightarrow{r'},\, t')$of observer $\mathcal{O}'$
in $R^{6}$ will appear as $T^{4}\equiv(t_{x}',\, t_{y}',\, t_{z}',\, r')$
to the observer $\mathcal{O}$ in $R^{6}$ The resulting space for
bradyons and tachyons is thus identified as the $R^{6}$ - or $M(3,\,3)$
space where both space and time and hence energy and momentum are
considered as vector quantities. Superluminal Lorentz transformations
(SLTs) between two frames $K$ and $K'$ moving with velocity $v>1$(in
the natural units $c=\hbar=1$) are defined in $R^{6}$ - or $M(3,\,3)$
space as follows;

\begin{eqnarray}
x' & \rightarrow & \pm t_{x};\nonumber \\
y' & \rightarrow & \pm t_{y};\nonumber \\
z' & \rightarrow & \pm\gamma(z-vt_{z});\nonumber \\
t_{x}' & \rightarrow & \pm x;\nonumber \\
t_{y}' & \rightarrow & \pm y;\nonumber \\
t_{z}' & \rightarrow & \pm\gamma(t_{z}-vz),\label{eq:1}\end{eqnarray}
where $\gamma=(v^{2}-1)^{\frac{1}{2}}$. The above transformations
lead to the mixing of space and time coordinates for transcendental
tachyonic objects, $(|\overrightarrow{v}|\rightarrow\infty)$ where
equation (\ref{eq:1}) takes the following form;

\begin{eqnarray}
+\, dt_{x}\, & \rightarrow dt_{x}'\,= & dx\,\,\,\,+\nonumber \\
+\, dt_{y}\, & \rightarrow dt_{y}'\,= & dy\,\,\,\,+\nonumber \\
+\, dt_{z}\, & \rightarrow dt_{z}'\,= & dz\,\,\,\,+\nonumber \\
-\, dz\, & \rightarrow dz'\,= & dt_{z}\,\,\,\,-\nonumber \\
-\, dy\, & \rightarrow dy'\,= & dt_{y}\,\,\,\,-\nonumber \\
-\, dx\, & \rightarrow dx'\,= & dt_{x}\,\,\,\,-.\label{eq:2}\end{eqnarray}
The above equation shows that we have only two four dimensional slices
of $R^{6}$ - or $M(3,\,3)$ space $(+,+.+,-)$ and $(-,-,-,+)$ .
When any reference frame describes bradyonic objects it is necessary
to describe $M(1,3)=[t,\, x,\, y,\, z]\,\,\,(R^{4}-\, space)$ so
that the coordinates $t_{x}$ and $t_{y}$ are not observed or couple
together giving $t=|\overrightarrow{t}\,|\equiv|(t_{x}^{2}+t_{y}^{2}+t_{z}^{2})|^{\frac{1}{2}}$.
On the other hand when a frame describes bradyonic object in frame
K, it will describe a tachyonic object (with velocity $(|\overrightarrow{v}|\rightarrow\infty)$
in $K'$ with $M'(1,\,3)$ space i.e. $M'(1,3)=[(t_{x}',\, x',\, y',\, z']=[z,\, t_{x},\, t_{y},\, t_{z}]\,\,\,(T^{4}-\, space)$.
We define $M'(1,\,3)$ space as $T^{4}$ - space or $M(3,\,1)$ space
where $x$ and $y$ are not observed or coupled together giving rise
to $|\overrightarrow{r\,}|=(x^{2}+y^{2}+z^{2})^{\frac{1}{2}}$. As
such, the spaces $R^{4}$ and $T^{4}$ are two observational slices
of $R^{6}$ - or $M(3,\,3)$ space but unfortunately the space is
not consistent with special theory of relativity. Subluminal and superluminal
Lorentz transformations lose their meaning in $R^{6}$- or $M(3,\,3)$
space with the sense that these transformations do not represent either
the bradyonic or tachyonic objects in this space. It has been shown
earlier (\cite{key-8,key-9} that the true localizations space for
bradyons is $R^{4}$ - space while that for tachyons is $T^{4}$ -
space. So a bradyonic $R^{4}=M(1,\,3)$ space maps to a tachyonic
$T^{4}=M'(3,\,1)$ space or vice versa i.e.

\begin{eqnarray}
R^{4} & =M(1,\,3)\overset{SLT}{\rightarrow} & M'(3,\,1)=T^{4}\label{eq:3}\end{eqnarray}
Let us describe the supersymmetric quantum mechanics (\cite{key-11})
in terms of a pair of bosonic Hamiltonians $H_{+}^{T}$ and $H_{-}^{T}$
which are supersymmetric partners (\cite{key-12}) of supersymmetric
Hamiltonian for tachyons in $T^{4}$- space i.e.,

\begin{eqnarray}
H^{T} & = & H_{+}^{T}\oplus H_{+}^{T}.\label{eq:4}\end{eqnarray}
In order to construct these super partner Hamiltonians for the system
described by $H^{T}=H_{B}^{T}\oplus H_{F}^{T}$ , we may introduce
the potential $V_{-}(t)$ whose ground state energy has been adjusted
to zero with the corresponding ground state wave function $\psi_{0}^{T(-)}$
given by;

\begin{eqnarray}
\psi_{0}^{T(-)} & = & \exp[-\int_{0}^{t}W(t')dt'].\label{eq:5}\end{eqnarray}
Substituting it in the Schrödinger equation (in the units of $\hbar=2k=1$
), we get

\begin{eqnarray}
V_{-}(t) & = & W^{2}(t)-W'(t)=\frac{\psi_{0}^{T''(-)}}{\psi_{0}^{(-)}}.\label{eq:6}\end{eqnarray}
Correspondingly for the potential, we have the following Hamiltonian

\begin{eqnarray}
H_{-}^{T} & =-\frac{d^{2}}{dt^{2}}+\frac{\psi_{0}^{T''(-)}}{\psi_{0}^{(-)}} & =-\frac{d^{2}}{dt^{2}}+V_{-}(t).\label{eq:7}\end{eqnarray}
If the ground state wave function $\psi_{0}^{T(-)}$ is square integrable
then the supersymmetry will be considered to be broken. This Hamiltonian
may also be written in the following form in terms of bosonic operator
$\hat{B}$ and $\hat{B}'$;

\begin{eqnarray}
H_{-}^{T} & = & \hat{B}^{+}\hat{B}\label{eq:8}\end{eqnarray}
where

\begin{eqnarray}
\hat{B} & =\frac{d}{dt}+W(t) & =\frac{d}{dt}-\frac{\psi_{0}^{T''(-)}}{\psi_{0}^{(-)}};\label{eq:9}\end{eqnarray}

\begin{eqnarray}
\hat{B}^{+} & =-\frac{d}{dt}+W(t) & =-\frac{d}{dt}-\frac{\psi_{0}^{T''(-)}}{\psi_{0}^{(-)}}.\label{eq:10}\end{eqnarray}
Let us introduce the Hamiltonian

\begin{eqnarray}
H_{+}^{T} & =\hat{B}\hat{B}^{+}=-\frac{d^{2}}{dt^{2}}+V_{+}(t) & =-\frac{d^{2}}{dt^{2}}+W^{2}(t)+W'(t)\label{eq:11}\end{eqnarray}
where

\begin{eqnarray}
V_{+}(t) & =W^{2}(t)+W'(t) & =V_{-}(t)+2W'(t).\label{eq:12}\end{eqnarray}
The potentials $V_{+}(t)$ and $V_{-}(t)$ are called supersymmetric
partner potentials and $H_{+}$ is the Hamiltonian corresponding to
the potential $V_{+}(t)$ . We may now impose the following condition
on the supersymmetry partner potentials $V_{+}(t)$ and $V_{-}(t)$
as ;

\begin{eqnarray}
V_{+}(t;c_{0}) & = & V_{-}(t;c_{1})+R(c_{1})\label{eq:13}\end{eqnarray}
where $c_{0}$ is a set of parameters occurring in $V_{+}(t)$ and
$c_{1}$ is a function of $c_{0}$ while the remainder $R(c_{1})$
is independent of $t$ . Then the supersymmetry partner potentials
$V_{+}(t)$ and $V_{+}(t)$ are said to be shape invariant. All potentials,
which exhibit the property of shape invariance, can exactly be solved.
It can be demonstrated in the straightforward manner by constructing
the sequence $H^{T(k)}$ of the Hamiltonians as 

\begin{eqnarray}
H^{T(k)} & = & -\frac{d^{2}}{dt^{2}}+V_{-}(t;c_{k})+\sum_{i=1}^{k}R(c_{i})\label{eq:14}\end{eqnarray}
where $c_{1}=f(c_{0})\,\, and\,\, c_{i}=f^{i}(c_{0})=f(c_{0}).....i\, times.$
It is obvious that

\begin{eqnarray}
H^{T(0)} & = & H^{T(1)},\,\,\,\, H_{-}^{T}=H_{+}^{T}.\label{eq:15}\end{eqnarray}
For all $k>o$, $H^{T(k)}$ and $H^{T(k+1)}$ are supersymmetry partner
Hamiltonians since they have identical bound state spectra except
for the lowest level of $H^{T(k)}$ i.e.;

\begin{eqnarray}
E_{0}^{T(k)} & = & \sum_{i=1}^{k}R(c_{i}).\label{eq:16}\end{eqnarray}
Furthermore, the ground state energy of $H^{T(k+1)}$ coincides with
the first excited state energy of $H^{T(k)}$ for all $k>o$ . Thus
we have ground state energy $H^{T(n)}$ as the $n^{th}$ state energy
of $H^{T(0)}$ . Combining this result with equation (\ref{eq:16}),
we get

\begin{eqnarray}
E_{0}^{T(k)} & = & \sum_{i=1}^{k}R(c_{i}),\,\, E_{0}^{T(-)}=0.\label{eq:17}\end{eqnarray}
For the Harmonic oscillator, we may define

\begin{eqnarray}
V_{+}(t) & = & V_{-}(t)+2\Omega.\label{eq:18}\end{eqnarray}
Here $\Omega$ is considered as the part of one - dimensional supersymmetric
harmonic oscillator $W(t)=\Omega t$ and with $H_{-}^{T}=\frac{d^{2}}{dt^{2}}+\Omega^{2}t^{2}-\Omega$
with $V_{-}(t)=\Omega^{2}t^{2}-\Omega$ . Comparing equation (\ref{eq:18})
with equation (\ref{eq:13}), we get

\begin{eqnarray}
c_{0}= & c_{1}=\Omega,\,\,\, & R(c_{1})=2\Omega.\label{eq:19}\end{eqnarray}
Then equation (\ref{eq:14}) becomes

\begin{eqnarray}
H^{T(k)} & = & \frac{d^{2}}{dt^{2}}+\Omega^{2}t^{2}+(2k-1)\Omega\label{eq:20}\end{eqnarray}
which reproduces the results (\ref{eq:15}). Equation (\ref{eq:20})
may also be written as

\begin{eqnarray}
H^{T(k+1)} & =\frac{d^{2}}{dt^{2}}+\Omega^{2}t^{2}+(2k+1)\Omega & =H^{T(k)}+\Omega'.\label{eq:21}\end{eqnarray}
Then equation (\ref{eq:16}) gives

\begin{eqnarray}
E_{0}^{T(n)} & = & n\Omega'\label{eq:22}\end{eqnarray}
showing that the ground state energy of $H^{T(n)}$ is identical with
the $n^{th}$ state energy of $H^{T(0)}=H_{-}^{T}$ . In the similar
manner, using the shape invariance property of partner potentials
$V_{+}(t)$ and $V_{-}(t)$ , since for shape invariant potential
we have may reproduce the other results of supersymmetric harmonic
oscillator

\begin{eqnarray}
\psi_{n}^{T(+)}(t;c_{0}) & = & \psi_{n}^{T(-)}(t;c_{1}).\label{eq:23}\end{eqnarray}
The method of shape invariant potentials used here for harmonic oscillator
can thus be generalized to all shape invariant potentials for the
deeper understanding of analytically solvable potentials. It is obvious
from equation (\ref{eq:21}) that operator $H_{+}^{T}=BB^{+}$is positive
definite for all states while the operator $H_{-}^{T}=B^{+}B$ is
semi - positive definite since $(B^{+}B)^{-1}$ is singular for $n=0$
in equation (\ref{eq:22}). Super partner Hamiltonians $H_{-}^{T}$
and $H_{+}^{T}$ are obviously Hermitian. Thus we can construct the
following operators

\begin{eqnarray}
U=H_{+}^{T(-\frac{1}{2})}B,\,\,\,\, & U^{+} & =B^{+}H_{+}^{T(-\frac{1}{2})}\label{eq:24}\end{eqnarray}
which gives

\begin{eqnarray}
UU^{+}=I,\,\,\,\, & U^{+}U & =B^{+}H_{+}^{T(-)}B=P\neq I\label{eq:25}\end{eqnarray}
showing that the operator $U$ is semi unitary and hence any transformation
involving $U$ will be semi unitary transformation (SUT). The operator
$P$ defined by equation (\ref{eq:25}) satisfies the following conditions

\begin{eqnarray}
P^{2} & =P,\,\,\, & P^{+}=P\label{eq:26}\end{eqnarray}
showing that it is a projection operator having the eigen values $0$
and $1$ . Under the SUT defined by equation (\ref{eq:24}), we have

\begin{eqnarray}
UH_{-}^{T}U^{+} & = & H_{+}^{T}.\label{eq:27}\end{eqnarray}
We also have

\begin{eqnarray}
[P,H_{-}^{T}] & = & 0\label{eq:28}\end{eqnarray}
which shows that $H_{-}^{T}$ and $P$ have common eigen functions
and eigen values of $P$ are good quantum members in $T^{4}$- space.
Using equation

\begin{eqnarray}
\hat{B^{+}\hat{B}}\psi^{T(-)}=E^{T}\psi^{T(-)},\,\,\, & \hat{B}\hat{B^{+}\hat{(B}}\psi^{T(-)})=E^{T}\hat{B}\psi^{T(-)},\,\,\, & H_{+}^{T}\hat{[B}\psi^{T(-)}]=E^{T}\hat{[B}\psi^{T(-)}]\label{eq:29}\end{eqnarray}
and $E_{n}^{T(+)}=E_{n+1}^{T(-)}$, we get

\begin{eqnarray}
P\psi_{n}^{T(-)} & =\psi_{n}^{T(-)}\,\,\, for\,\, n>0,\,\,\, & P\psi_{0}^{T(-)}=0.\label{eq:30}\end{eqnarray}
In general we know

\begin{eqnarray}
P\psi_{n+1}^{T(-)} & = & \psi_{n+1}^{T(-)}.\label{eq:31}\end{eqnarray}
Let us denote

\begin{eqnarray}
\psi_{0}^{T(-)}=|0,0>,\,\,\,\, & \psi_{n+1}^{T(-)} & =|n+1,1>\label{eq:32}\end{eqnarray}
as the eigen state for harmonic oscillator, we may readily obtain,

\begin{eqnarray}
H_{-}^{T}|n+1,1> & = & (n+1)\Omega|n+1,1>\nonumber \\
P|n+1,1> & = & |n+1,1>\nonumber \\
H_{-}^{T}|0,0> & = & 0;\nonumber \\
P|0,0> & = & 0.\label{eq:33}\end{eqnarray}
It shows that under the projection $P$ the full Hilbert space $H$
of harmonic oscillator is projected in two subspaces

\begin{eqnarray}
H & = & H_{0}\bigoplus H_{1}\label{eq:34}\end{eqnarray}
where the subspace $H_{0}$ is constituted by the state $|0,0>$ and
the subspace $H_{1}$ is constituted by the states $|n+1,1>$ . To
find the general structure of the operator $P$ in the above basis,
let us start with the complete set $|n>$ of $H_{-}^{T}$ (for $n=0,1,2---$
), and construct

\begin{eqnarray}
|n>_{+} & =U^{+}|n> & =H_{+}^{T(-\frac{1}{2})}B|n>\label{eq:35}\end{eqnarray}
which gives

\begin{eqnarray}
\sum_{n=0}^{\infty}|n>_{++}<n| & =U^{+}\sum_{n=0}^{\infty}|n><n|U & =U^{+}U=P\label{eq:36}\end{eqnarray}
showing that states $|n>_{+}$do not form the complete set. We obviously
have

\begin{eqnarray}
P|n>_{+}=|n>_{+},\,\,\,\, & |n>_{+} & =|P=1>.\label{eq:37}\end{eqnarray}
The complete set of the eigen states is thus formed by the states
$|q=0>$ and $|n>_{+}$ i.e.

\begin{eqnarray}
|q=0><q=0|+\sum_{n}|n>_{++}<n|=I,\,\,\,\, & \sum_{n}|n>_{++}<n| & =I-|0><0|.\label{eq:38}\end{eqnarray}
As such, the matrix of the operator $P$ is diagonal in the basis
given by equation (\ref{eq:31}) with the general structure

\begin{eqnarray}
P & =\left(\begin{array}{ccccc}
0 & 0 & 0 & ... & 0\\
0 & 1 & 0 & ... & 0\\
0 & 0 & 1 & ... & 0\\
\vdots & \vdots & \vdots & ... & \vdots\\
0 & 0 & 0 & ... & 1\end{array}\right) & =I-|0><0|.\label{eq:39}\end{eqnarray}
Furthermore, from SUT transformation (\ref{eq:35}) we get

\begin{eqnarray}
_{+}<n,n>_{+} & =<n|UU^{+}|n> & =<n,n>\label{eq:40}\end{eqnarray}
showing that the orthonormality of the states $|n>$ implies the orthonormality
of states $|n>_{+}$. Thus equations (\ref{eq:38}) and (\ref{eq:40})
demonstrate that under the SUT transformation (\ref{eq:35}) the orthogonality
and the normalization of states are maintained while the completeness
condition is violated. On the other hand, let us consider the SUT
transformation

\begin{eqnarray}
|n>_{-} & = & U|n>\label{eq:41}\end{eqnarray}
where $|n>$constitute the complete orthonormal set of states of $H_{-}^{T}$
. Then we have

\begin{eqnarray}
\sum_{n=0}^{\infty}|n>_{--}<n| & =\sum_{n=0}^{\infty}U|n><n|U^{+} & =UU^{+}=I\label{eq:42}\end{eqnarray}
showing that the completeness of set $|n>$ implies the completeness
of set $|n>_{-}$. But

\begin{eqnarray}
_{-}<m,n>_{-}=<m|U^{+}U|n> & =<m|P|n> & \neq\delta_{mn}.\label{eq:43}\end{eqnarray}
In other words the transformation (\ref{eq:41}) maintains completeness
relation but fails to maintain orthogonality and normalization conditions.
However, under the transformation (\ref{eq:35}) we have

\begin{eqnarray}
|0,0>_{-} & =U|0,0> & =0\label{eq:44}\end{eqnarray}
which shows that this SUT destroys the ground state $|0,0>$ , while
the state

\begin{eqnarray}
|n+1,1>_{-} & = & U|n>\label{eq:45}\end{eqnarray}
satisfies the following conditions by using relation (\ref{eq:44})

\begin{eqnarray}
\sum_{n}|n+1,1>_{--}<n+1,1| & =UU^{+} & =I\,\,\,;\label{eq:46}\\
_{-}<m+1,1|n+1,1>_{-} & =<m+1|P|n+1> & =\nonumber \\
<m+1|I-0,0><0,0|n+1> & = & <m+1|n+1>=\delta_{mn}.\label{eq:47}\end{eqnarray}
Thus the orthogonality and normalization conditions are restored for
the states $|n+1,1>_{-}$ , which constitute the basis of $H_{1}$
. In other words, the states $|n+1,1>_{+}$ constitute complete orthonormal
set, even though $|0,0>_{-}$ , is destroyed. It shows that the supersymmetric
partner $H_{+}^{T}$, compared with $H_{-}^{T}$ , loses the ground
state but its eigen function $\psi_{n}^{T(+)}=U\psi_{n+1}^{T(-)}$
with eigen values $E_{n}^{T(+)}=E_{n+1}^{T(-)}$ constitutes a complete
orthonormal set in the subspace $H_{1}$ . Thus the SUT introduced
here provides a new way to relate solvable systems to their supersymmetric
partners and helps us to construct a new class of solvable potentials
when we start from solvable systems where Hamiltonian can be factorized.
It has shown that semi unitary transformations operators $U$ and
$U^{+}$ constructed above describes a projection operator having
eigen values $0$ and $1$ while the commutation relation shows that
$H_{-}^{T}$ and $P$ have common eigen functions and eigen values
in $T^{4}$ - space. It has been discussed that under the projection
$P$ the full Hilbert space of harmonic oscillator is decomposed in
two subspaces and the states $|n>_{+}$ associated with semi unitary
operator in $T^{4}$ - space form the complete set. The diagonalization
of projection operator $P$ has been illustrated and accordingly the
orthonormality condition leads to the conclusion that the orthonormality
of state $|n>$ implies the orthonormality of states $|n>_{+}$. As
such it is claimed that under the SUT transformation the orthogonality
and normalization of states are maintained while the completeness
condition has been said to be violated. It has also been shown that
the orthogonality and normalization conditions are restored for the
states $|n+1,1>_{-}$. In other words the states $|n+1,1>_{-}$ constitute
complete orthonormal set, even though $|0,0>_{-}$ is destroyed. It
shows that supersymmetric partner $H_{+}^{T}$ , compared with $H_{-}^{T}$
, loses the ground state but its eigen function $\psi_{n}^{T(+)}=U\psi_{n+1}^{T(-)}$
with eigen value $E_{n}^{T(+)}=E_{n+1}^{T(-)}$ constitute a complete
orthonormal set. It has already been emphasized earlier \cite{key-8,key-9}
that on passing from bradyons to tachyons, the role of space and time
and consequently momentum and energy are changed. Thus the positivity
of energy for bradyonic particles is considered only in four - dimensional
space with three space and one time coordinates i.e. in $R^{4}$ -
space while for the case of tachyons the Hamiltonian is space dependent
in $T^{4}$ - space and the positivity of momentum is being taken
in to account. The $T^{4}$ - space has been visualized as the space
for tachyons where they behave as bradyons do in $R^{4}$ - space.
The fore going analysis for tachyons and their behaviour in supersymmetric
quantum mechanics leads to the conclusion that there has been basic
disagreement in localization and representation of tachyons. As such
the observables of more than four dimensions of space - time may be
associated with the even horizon effects ($R^{4}\rightarrow T^{4}$)
taking into account the interconnection between boson - fermion symmetry
and bradyon tachyon transformations. The semi unitary transformation
(SUT) introduced here provides a new window to relate solvable system
to their supersymmetric partners and helps us to construct a new class
of solvable potentials when we start from solvable system whose Hamiltonian
can be factorized on passing from bradyons to tachyons in $T^{4}$
- space.

\textbf{Acknowledgement:} This work is carried out under Uttarakhand
State Council for Science and Technology, Dehradun research scheme.

\end{document}